\documentclass[a4paper,twocolumn,english,pra,eqseqnum]{revtex4}
\usepackage[T1]{fontenc}
\usepackage[latin1]{inputenc}
\usepackage{graphicx}
\usepackage{amssymb}

\makeatletter

\date{\today}
\usepackage{ae,aecompl}

\usepackage{bm}

\usepackage{babel}
\makeatother
\begin{document}

\title{Gaussian quantum Monte Carlo methods for fermions}

\author{J. F. Corney and P. D. Drummond}

\affiliation{ARC Centre of Excellence for Quantum-Atom Optics, University of Queensland,
Brisbane 4072, Queensland, Australia.}

\begin{abstract}
We introduce a new class of quantum Monte Carlo methods, based on
a Gaussian quantum operator representation of fermionic states. The
methods enable first-principles dynamical or equilibrium calculations
in many-body Fermi systems, and, combined with the existing Gaussian
representation for bosons, provide a unified method of simulating
Bose-Fermi systems. As an application, we calculate finite-temperature
properties of the two dimensional Hubbard model.
\end{abstract}
\maketitle
Calculating the quantum many-body physics of interacting Fermi systems
is one of the great challenges in modern theoretical physics. These
issues appear in physical problems at all energy scales, from ultra-cold
atomic physics to high-energy lattice QCD. In even the simplest cases,
first-principles calculations are inhibited by the complexity of the
fermionic wavefunction, manifest notoriously in the Fermi sign problem.
In previous quantum Monte Carlo (QMC) techniques, the sign problem
appears as trajectories with negative weights, which contribute to
a large sampling error\cite{Ceperley99}. QMC methods are also complicated
by the calculation of large determinants.

In this letter, we introduce a new QMC method for simulating many-body
fermion systems, based on a Gaussian phase-space representation. As
an application to condensed matter and AMO physics, we study the well-known
Hubbard model. Although it is the simplest model of interacting fermions
on a lattice, it is rich in physics and may even describe high-temperature
superconductivity\cite{Linden92}. We show that for the Hubbard model
the Gaussian representation leads to imaginary-time equations with
no negative probabilities or weights. We demonstrate that this removes
the well-known Fermi sign problem\cite{Ceperley99,Linden92,Santos03},
by first principles numerical simulation without fixed-node\cite{Astrakharchik_etal}
or variational approximations.

Phase-space methods\cite{ICOLSproceedings} provide a way to simulate
quantum many-body systems both dynamically and at finite temperature,
and have proved useful in bosonic cases. These methods sample the
time evolution of a positive distribution on an overcomplete basis
set, which is usually the set of coherent states. However, whereas
coherent state representations are well-defined in the bosonic case,
the only known coherent state techniques for fermions involve Grassmann
algebra\cite{Grassmann}, which has an enormous computational complexity. 

Here we introduce a phase-space method that overcomes the problem
of Grassmann complexity, using a Gaussian expansion for fermions.
The operator basis is constructed from pairs \emph{}of Fermi operators.
Because these pairs obey commutation relations, a natural solution
of the Grassmann problem is achieved. Furthermore, the resulting equations
obviate the need to evaluate large determinants. The elimination of
anti-commutators means that the technique is far more efficient than
previous QMC and stochastic fermion methods\cite{Stochastic_Fermi_methods}.
We give examples in cases of experimental relevance involving the
dynamical problem of Pauli blocking in molecular dissociation, and
finite temperature correlations of fermions in an optical lattice,
where the results agree with those of other exact methods. We also
perform larger simulations of the 2D Hubbard model, in cases where
severe sign problems were found previously.

Our starting point is a general expansion of the system density operator:

\begin{equation}
\widehat{\rho}(t)=\int P(\overrightarrow{\lambda},t)\widehat{\Lambda}(\overrightarrow{\lambda})d\overrightarrow{\lambda}\,\,,\label{eq:general-expansion}\end{equation}
where $P(\overrightarrow{\lambda},t)$ is a probability distribution,
$\widehat{\Lambda}$ is a suitable basis for the class of density
matrices being considered, and $d\overrightarrow{\lambda}$ is the
integration measure for the corresponding generalized phase-space
coordinate $\overrightarrow{\lambda}$. The operators $\widehat{\Lambda}$
are non-Hermitian and form a complete basis for density operator.
Existing phase-space methods are defined for systems of bosons, with
$\widehat{\Lambda}$ constructed of bosonic ladder operators\cite{Gauss:Bosons}. 

To achieve a unified representation, we define the operator basis
$\widehat{\Lambda}$ to be the product of Gaussian forms of bosonic
and fermionic creation and annihilation operators: $\widehat{\Lambda}\equiv\Omega\widehat{\Lambda}_{b}\widehat{\Lambda}_{f}$,
where $\widehat{\Lambda}_{b}$ and $\widehat{\Lambda}_{f}$ are Gaussian
forms over $M_{b}$ bosonic modes and $M$ fermionic modes, respectively,
and where the (possibly) complex number $\Omega$ is an additional
weighting factor. The properties of the bosonic Gaussian representation
are given in \cite{Gauss:Bosons}. Here we summarise the relevant
properties of the Fermionic Gaussian form; explicit proofs will be
given elsewhere.

For a system that can be decomposed into $M$ single-particle modes,
we define $\widehat{\bm a}$ as a column vector of the $M$ annihilation
operators, and $\widehat{\bm a}^{\dagger}$ as the corresponding row
vector of $M$ creation operators, whose anticommutation relations:
$[\widehat{a}_{k},\widehat{a}_{j}^{\dagger}]_{+}=\delta_{kj}\,\,.$
We also introduce an extended $2M$-vector of all the operators: $\underline{\widehat{a}}=(\widehat{\bm a},(\widehat{\bm a}^{\dagger})^{T})\,\,$,
with adjoint defined as $\underline{\widehat{a}}^{\dagger}=(\widehat{\bm a}^{\dagger},\widehat{\bm a}^{T})\,\,$.
A general, normally ordered Gaussian operator can then be written

\begin{eqnarray}
\widehat{\Lambda}_{f} & = & \mathrm{Pf}\left[\underline{\underline{\sigma_{A}}}\right]:\exp\left[-\underline{\widehat{a}}^{\dagger}\left(\underline{\underline{I}}-\underline{\underline{\sigma}}^{-1}/2\right)\underline{\widehat{a}}\right]:\,\,,\label{eq:Gaussbasis}\end{eqnarray}
 which, because it is constructed from pairs of operators, contain
no Grassmann variables. The normalisation, chosen to ensure that ${\rm Tr}\,\widehat{\Lambda}_{f}=1\,$,
consists of the Pfaffian of an antisymmetric form $\underline{\underline{\sigma_{A}}}$
of the covariance\cite{Pfaffian}.

Normal ordering, denoted by $:\cdots:$ , is defined as in the bosonic
case, with all annihilation operators to the right of the creation
operators, except that each pairwise reordering involved induces a
sign change, e.~g.~$:\widehat{a}_{i}\widehat{a}_{j}^{\dagger}:\,=-\widehat{a}_{j}^{\dagger}\widehat{a}_{i}\,\,$.
We define \emph{anti}normal ordering similarly, and denote it via
curly braces: $\{\widehat{a}_{j}^{\dagger}\widehat{a}_{i}\}=-\widehat{a}_{i}\widehat{a}_{j}^{\dagger}\,\,$.
More generally, we can define nested orderings, in which the outer
ordering does not reorder the inner one. For example, $\{:\widehat{\Lambda}\widehat{a}_{j}^{\dagger}:\widehat{a}_{i}\}=-\widehat{a}_{i}\widehat{a}_{j}^{\dagger}\widehat{\Lambda}\,\,$,
where we assume that the kernel $\widehat{\Lambda}$ always remains
normally ordered.

The generalized covariance $\underline{\underline{\sigma}}$ and constant
matrix $\underline{\underline{I}}$ are $2M\times2M$ matrices, which
we can write as

\begin{eqnarray}
\underline{\underline{\sigma}}=\left[\begin{array}{cc}
\mathbf{I}-\mathbf{n}^{T} & \mathbf{-m}\\
\mathbf{-m}^{+} & \mathbf{n}-\mathbf{I}\end{array}\right] & ,\, & \underline{\underline{I}}=\left[\begin{array}{cc}
\bm I & \bm0\\
\bm0 & -\bm I\end{array}\right]\,\,,\label{eq:covariance}\end{eqnarray}
where the number correlation $\mathbf{n}$ is a complex $M\times M$
matrix, the squeezing correlations $\mathbf{m},\,\mathbf{m}^{+}$
are two independent antisymmetric complex $M\times M$ matrices, and
$\bm I$ is the $M$-mode identity matrix. 

The phase space of the fermionic representation is $\overrightarrow{\lambda}=(\Omega,\mathbf{n},\mathbf{m},\mathbf{m}^{+})$,
which has a dimension of $1+p=1+M(2M-1)\,\,$. For a combined Bose-Fermi
system, there will be an additional $M_{b}(2M_{b}+3)$ bosonic dimensions.

Under the Gaussian representation, physical quantities (operator expectation
values) appear as moments of the (weighted) distribution $\Omega P$,
denoted as $\left\langle ..\right\rangle _{P}$. For quadratic products,

\begin{eqnarray}
\left\langle \widehat{a}_{i}\widehat{a}_{j}\right\rangle  & = & \left\langle m_{ij}\right\rangle _{P}\,\,,\nonumber \\
\left\langle \widehat{a}_{i}^{\dagger}\widehat{a}_{j}^{\dagger}\right\rangle  & = & \left\langle m_{ij}^{+}\right\rangle _{P}\,\,,\nonumber \\
\left\langle \widehat{a}_{i}^{\dagger}\widehat{a}_{j}\right\rangle  & = & \left\langle n_{ij}\right\rangle _{P}\,\,.\label{eq:moments}\end{eqnarray}
For higher order products, the corresponding moments can be determined
by evaluation of the appropriate (Grassmann) Gaussian integral. Note
that there is no way to calculate the expectation value of single
ladder operators (or any product that is of odd order); the Gaussian
form cannot represent density operators that contain an odd number
of operators. However, in physical Hamiltonians, Fermi operators appear
only in pairs, and so such `odd' states will not be generated in the
course of the evolution. For other, physical states, the Gaussian
basis provides an (over)complete representation. 

To enable canonical or dynamical simulations, we need identities that
describe the action of operators on the density operator as derivatives
on elements of the Gaussian basis. With our ordering notation given
above, all of the necessary operator identities can be written in
a compact matrix form:

\begin{eqnarray}
\widehat{\Lambda} & = & \Omega\frac{\partial}{\partial\Omega}\widehat{\Lambda}\,\,,\nonumber \\
:\widehat{\underline{a}}\,\widehat{\underline{a}}^{\dagger}\widehat{\Lambda}: & = & -\underline{\underline{\sigma}}\widehat{\Lambda}+\underline{\underline{\sigma}}\frac{\partial\widehat{\Lambda}}{\partial\underline{\underline{\sigma}}}\underline{\underline{\sigma}}\,\,,\nonumber \\
\left\{ \widehat{\underline{a}}:\widehat{\underline{a}}^{\dagger}\widehat{\Lambda}:\right\}  & = & \underline{\underline{\sigma}}\widehat{\Lambda}-\left(\underline{\underline{\sigma}}-\underline{\underline{I}}\right)\frac{\partial\widehat{\Lambda}}{\partial\underline{\underline{\sigma}}}\underline{\underline{\sigma}}\,\,,\nonumber \\
\left\{ \widehat{\underline{a}}\,\widehat{\underline{a}}^{\dagger}\widehat{\Lambda}\right\}  & = & -\left(\underline{\underline{\sigma}}-\underline{\underline{I}}\right)\widehat{\Lambda}+\left(\underline{\underline{\sigma}}-\underline{\underline{I}}\right)\frac{\partial\widehat{\Lambda}}{\partial\underline{\underline{\sigma}}}\left(\underline{\underline{\sigma}}-\underline{\underline{I}}\right).\label{eq:Matrixidentities}\end{eqnarray}
 The matrix derivative is here defined as $(\partial/\partial\underline{\underline{\sigma}})_{\mu,\nu}=\partial/\partial\sigma_{\nu\mu}\,\,$.
Notice that there are no determinants (or Pfaffians) to be calculated
in these identities. 

As an application of the fermionic representation, consider the well-known
Hubbard model: $H(\widehat{\mathbf{n}}_{\upharpoonleft},\widehat{\mathbf{n}}_{\downharpoonleft})=$\begin{equation}
-t\sum_{\left\langle i,j\right\rangle ,\sigma}\widehat{n}_{i,j.\sigma}+U\sum_{j}\widehat{n}_{j,j,\upharpoonleft}\widehat{n}_{j,j,\downharpoonleft}-\mu\sum_{j,\sigma}\widehat{n}_{j,j,\sigma}\,,\label{eq:Hubbard}\end{equation}
 where $\widehat{n}_{i,j,\sigma}=\widehat{a}_{i,\sigma}^{\dagger}\widehat{a}_{j,\sigma}$=$\left\{ \widehat{\mathbf{n}}_{\sigma}\right\} _{ij}$,
$t$ is the hopping, or tunelling, strength, $U$ is strength of on-site
interactations and $\mu$ is the chemical potential, included to control
the total particle number. The index $\sigma$ denotes spin ($\upharpoonleft,\downharpoonleft$)
and the indices $i,j$ label lattice location, with $\left\langle i,j\right\rangle $
denoting a sum over nearest neighbours. The Hubbard model is the simplest
nontrivial model for strongly interacting electrons and is thus an
important system in condensed matter physics, with relevance to the
theory of high-temperature superconductors\cite{Linden92}. It also
describes an ultracold fermi gas in a optical lattice potential. The
physics of the model is not yet fully understood, and although there
are known solutions in the 1D case\cite{LiebWu68}, this is not so
for higher dimensions. 

The 2D problem in particular is an important testing ground for QMC
methods. Traditional methods are prone to sign problems in the repulsive
case ($U>0$) away from half filling. These are particularly severe
for large systems, higher dimensions, stronger interaction and open-shell
configurations\cite{Santos03,FettesMorgenstern00}. 

The equilibrium state at temperature $T=1/k_{B}\tau$ can be cast
into an `imaginary time' differential equation for the unnormalised
density operator: $d\widehat{\rho}/d\tau=-\frac{1}{2}[\widehat{H}\,,\widehat{\rho}]_{+}\,\,$.
We make use of the representation by expanding the density operator
in terms of the Gaussian operators and applying the identities in
Eq.~(\ref{eq:Matrixidentities}). After integrating by parts, we
arrive at an equation for the distribution function, which we can
sample numerically, by solving stochastic phase-space equations. Although
there is never any need to calculate determinants with these methods,
the sampling error typically grows in (imaginary) time unless a suitable
choice of `stochastic gauge' is made\cite{Gauge}, in which one exploits
the overcomplete nature of the basis to keep the distribution compact.
Stochastic gauges can also be used to eliminate boundary terms that
may arise in the partial integration step.

Before applying this procedure to the Hubbard model, we first rewrite
the interaction terms as\begin{eqnarray}
U:\widehat{n}_{j,j,\upharpoonleft}\widehat{n}_{j,j,\downharpoonleft}: & = & -|U|/2:\left(\widehat{n}_{j,j,\upharpoonleft}-s\widehat{n}_{j,j,\downharpoonleft}\right)^{2}:,\end{eqnarray}
where $s=U/|U|$. The extra terms here vanish because of the anticommuting
property of fermion operators, but they do lead to additional stochastic
terms. Such terms are examples of a new type of stochastic gauge,
and one that is unique to fermions: vanishing operator products can
be used to modify the stochastic behaviour of the phase-space equations
without affecting the averaged results. With this choice of terms,
we map the imaginary-time calculation onto a set of \emph{real} Stratonovich
stochastic equations, which, in matrix form, are\begin{eqnarray}
\frac{d\mathbf{n}_{\sigma}}{d\tau} & = & \frac{1}{2}\left\{ \left(\mathbf{I}-\mathbf{n}_{\sigma}\right)\bm\Delta_{\sigma}^{(1)}\!\mathbf{n}_{\sigma}+\mathbf{n}_{\sigma}\bm\Delta_{\sigma}^{(2)}\!\left(\mathbf{I}-\mathbf{n}_{\sigma}\right)\right\} .\end{eqnarray}
Here we have introduced the matrix: $\Delta_{i,j,\sigma}^{(r)}=$\[
t\delta_{\left\langle i,j\right\rangle }-\delta_{i,j}\left\{ |U|(sn_{j,j,-\sigma}-n_{j,j,\sigma}+\frac{1}{2})-\mu+f\xi_{j}^{(r)}\right\} ,\]
where $\delta_{\left\langle i,j\right\rangle }=1$ if the $i,j$ correspond
to nearest neighbour sites and is otherwise $0$, and where $f=-s$
for $\sigma=\downharpoonleft$ and 1 otherwise. The real Gaussian
noise $\xi_{j}^{(r)}(\tau)$ is defined by the correlations $\left\langle \xi_{j}^{(r)}(\tau)\,\xi_{j'}^{(r')}(\tau')\right\rangle =2|U|\delta(\tau-\tau')\delta_{j,j'}\delta_{r,r'}\,\,$.
The weights for each trajectory evolve as physically expected for
energy-weighted averages, with $d\Omega/d\tau=-\Omega H(\mathbf{n}_{\upharpoonleft},\mathbf{n}_{\downharpoonleft})$.
Because the equations for the phase-space variables $n_{i,j,\sigma}$
are all real, the weights will all remain positive, thereby avoiding
the traditional manifestation of the sign problem. 

\begin{figure}
\includegraphics[%
  width=7.5cm,
  keepaspectratio]{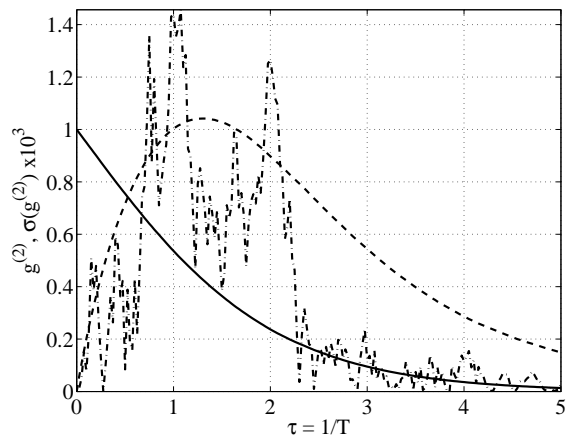}

\caption{\label{cap:Canonical-calculation}Second-order correlation function
$g^{(2)}$ versus inverse temperature $\tau$ for $t=0$, $U=2$ and
$\mu=1$, for which $\left\langle n_{j}\right\rangle =0.5$. The solid
curve gives the simulation result, and the dashed and dot-dashed line
show the estimated sampling error and deviation from the analytic
result, respectively (on a $\times1000$ scale). Calculated from 100,000
trajectories}
\end{figure}

Consider first the case where $t=0$, which describes, for example,
an ultracold Fermi gas in a deep optical lattice potential. In Fig.~\ref{cap:Canonical-calculation},
we plot the the correlation function $g^{(2)}\equiv\left\langle :\widehat{n}_{1}\widehat{n}_{2}:\right\rangle /\left\langle \widehat{n}_{1}\right\rangle \left\langle \widehat{n}_{2}\right\rangle $
for $U>0$, revealing a strong antibunching effect at low temperatures.
The sampling error here is very small, because of the restricted phase-space
explored by the simulation. 

\begin{figure}
\includegraphics[%
  width=7.5cm]{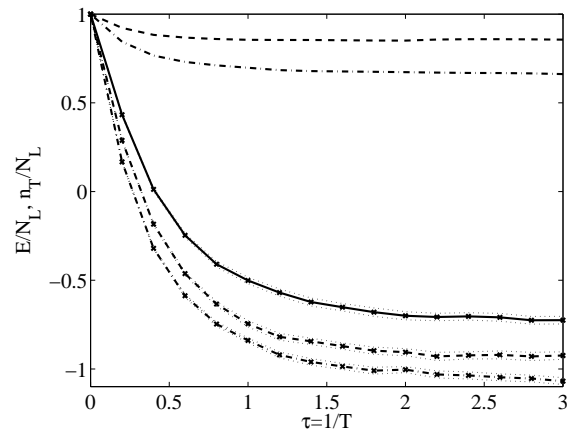}

\caption{\label{cap:Total-energy}Total energy $E$ per site versus inverse
temperature $\tau$ for a $16\times16$ 2D lattice for chemical potentials
$\mu=2$ (solid), $\mu=1$ (dashed) and $\mu=0$ (dot-dashed). Curves
without crosses give the number of particles per site for $\mu=1$
(dashed) and $\mu=0$ (dot-dashed). $U=4$, $t=1$, and 50 paths initially.
Dotted curves give an estimate of sampling error.}
\end{figure}

Whether the method can overcome the fundamental cause of the sign
problem, which is the complexity of fermionic states, must be demonstrated
by calculating physical quantities in cases where the sign problem
is known to occur in other methods. Thus we calculate the total energy
for $t=1$, $U=4$ in two dimensions as a function of temperature,
for different fillings. The results for a 16x16 lattice are shown
in Fig.~\ref{cap:Total-energy}, in which to obtain good sampling
with the spreading weights, we use a branching technique\cite{TrivediCeperley90}.
For a 4x4 lattice at an inverse temperature of $\tau=7$, we calculate
$E=-13.1\pm1.2$ at $n=0.5$ and $E=-19.62\pm0.87$ at $n=.313\pm0.005$,
which can be compared to zero-temperature, exact-diagonalisation results,
namely $E=-13.62$ for $n=0.5$ (half filling) and $E=-19.57$ for
$n=0.3125$ (10 atoms)\cite{Exact}. At a filling of $n=0.412\pm0.01$,
for which the sign deteriorates for a projector QMC calculation\cite{FettesMorgenstern00},
we calculate $E=-16.5\pm1.5$. We emphasise that, unlike projector
QMC \cite{Linden92}, the Gaussian method can calculate any physical
correlation function, at any temperature. 

\begin{figure}
\includegraphics[%
  width=7.5cm,
  keepaspectratio]{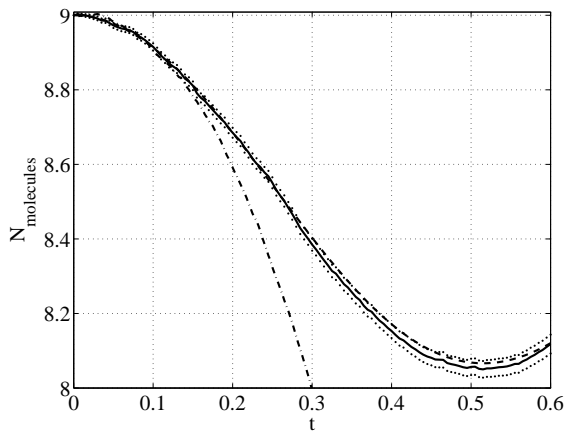}

\caption{\label{cap:Molecular-dissociation}Molecular dissociation into pairs
of fermionic (solid line) or bosonic (dot-dashed line) atoms. For
the fermionic case, the dashed curve gives the truncated number-state
calculation, and the dotted lines the estimated sampling error. In
the bosonic case, the estimated sampling error is too small to be
distinctly plotted on this graph. The initial state is a molecular
coherent state ( $N_{{\rm molecules}}(0)=9$). Calculated from 10,000
trajectories.}
\end{figure}

As an application of the method to a dynamical calculation, in particular
to a composite Bose/Fermi system, we consider the process of the dissociation
of a molecular Bose condensate into its constituent atoms, which may
be fermions or bosons. For simplicity, we consider two atomic modes,
representing, for example, states of different spin or momenta, coupled
to a single molecular mode, via the effective interaction $\hat{H}=\hat{a}^{\dagger}\hat{b}_{1}\hat{b}_{2}+\mathrm{h.c.}$,
where $\hat{b}_{j}^{\dagger}$, $\hat{b}_{j}$ are the atomic creation
and annihilation operators and $\hat{a}^{\dagger}$, $\hat{a}$ are
the bosonic molecular operators. Realistic models of the atomic-molecular
Feshbach resonances contain such terms to describe the coupling, and
it is important to illustrate how this method can represent them.
Because the normal spin-spin correlations $<\hat{b}_{1}^{\dagger}\hat{b}_{2}>$
will remain zero in this system (if initially zero), the phase space
of the system reduces to $\overrightarrow{\lambda}=(\alpha,\alpha^{+},n_{1,}n_{2},m,m^{+})\,\,,$
i.~e.~four complex atomic variables and two complex molecular amplitudes.
Applying the identities in Eq.~(\ref{eq:Matrixidentities}) and in
\cite{Gauss:Bosons}, we derive the following phase-space equations
for the time evolution, (where $+\rightarrow\mathrm{bosonic}$, $-\rightarrow\mathrm{fermionic})$:\begin{eqnarray}
\dot{n_{j}} & = & i(\alpha^{+}m-\alpha m^{+})\pm\sqrt{i}n_{j}\left(m\zeta_{1}^{\ast}+m^{+}\zeta_{2}^{\ast}\right)\,\,,\nonumber \\
\dot{m} & = & -i\alpha(1\pm n_{1}\pm n_{2})+\sqrt{i}\left(\pm m^{2}\zeta_{1}^{\ast}+n_{1}n_{2}\zeta_{2}^{\ast}\right)\,\,,\nonumber \\
\dot{m^{+}} & = & i\alpha^{+}(1\pm n_{1}\pm n_{2})+\sqrt{i}\left(n_{1}n_{2}\zeta_{1}^{\ast}\pm m^{+2}\zeta_{2}^{\ast}\right)\,\,,\nonumber \\
\dot{\alpha} & = & -im-\sqrt{i}\zeta_{1}\,\,,\,\,\dot{\alpha^{+}}=im^{+}+\sqrt{i}\zeta_{2}\,\,,\label{eq:am_phasespace}\end{eqnarray}
 where $j=1,2$ and where the $\zeta_{k}(t)$ are two complex Gaussian
noises, defined by the correlations $\left\langle \zeta_{k}(t)\zeta_{k'}(t')\right\rangle =0\,,\,\,\left\langle \zeta_{k}(t)\zeta_{k'}^{\ast}(t')\right\rangle =\delta_{k,k'}\delta(t-t')\,\,$.
Simulations of Eqs.~(\ref{eq:am_phasespace}) are compared with truncated
number-state-based calculations in Fig.~\ref{cap:Molecular-dissociation}.
Although the initial rates of conversion are the same in each case,
a Pauli blocking effect soon slows the fermionic conversion, in contrast
to an enhanced bosonic conversion. We note that in these real-time
calculations, a growing sampling error appears to be a generic property,
although a prudent gauge choice may control the growth rate for a
certain time.

In summary, we have introduced here an operator representation that
is able to represent arbitrary physical states of fermions. Together
with the corresponding bosonic representation, it is the largest class
of representations that can be constructed using an operator basis
that is Gaussian in the ladder operators. We have presented identities
for first-principles calculations of the time evolution of quantum
systems, both dynamical (real time) and canonical (imaginary time).
A quadratic master equation maps to deterministic equations, whereas
interacting systems with quartic terms in the Hamiltonian generate
stochastic equations, provided a suitable stochastic gauge is chosen
that eliminates all boundary terms. No computationally intensive determinant
calculations are involved. 

The simple examples given here show how the one, unified method can
solve both fermionic and bosonic problems, making it well suited to
simulating Bose-Fermi mixtures, and to studying from first-principles,
for example, the BEC/BCS crossover. Importantly, a new type of Fermi
stochastic freedom can be used to map canonical calculations of the
Hubbard type onto a real subspace. We have thereby been able to numerically
simulate the Hubbard model without sign error, even without employing
any of the sophisticated sampling techniques that have been developed
over time to optimise more conventional QMC methods. The application
of such techniques to the Gaussian approach is yet to be explored.

We gratefully acknowledge support from the Australian Research Council.


\begin{thebibliography}{10}
\bibitem{Ceperley99}D. M. Ceperley, Rev.~Mod.~Phys.~\textbf{71}, 438 (1999).
\bibitem{Linden92}W. von der Linden, Phys.~Rep.~220, 53 (1992).
\bibitem{Santos03}R. R. dos Santos, Brazilian Journal of Physics 33, 36 (2003). 
\bibitem{Astrakharchik_etal}G. E Astrakharchik \emph{et al}, cond-mat/0406113.  
\bibitem{ICOLSproceedings}P. D. Drummond, P. Deuar, J. F. Corney and K. Kheruntsyan, in \textit{Proceedings
of the 16th International Conference on Laser Spectroscopy}, ed. P.
Hannaford, A. Sidorov, H. Bachor and K. Baldwin (World Scientific,
2004), p 161.
\bibitem{Grassmann}K. E. Cahill and R. J. Glauber, Phys.~Rev.~A \textbf{59}, 1538 (1999);
L. I. Plimak, M. J. Collett and M. K. Olsen, Phys.~Rev.~A \textbf{64},
063409 (2001). 
\bibitem{Stochastic_Fermi_methods}O. Juillet, F. Gulminelli and Ph.~Chomaz, Phys.~Rev.~Lett.~\textbf{92},
160401 (2004).  
\bibitem{Gauss:Bosons}J. F. Corney and P. D. Drummond, Phys.~Rev.~A \textbf{68}, 063822
(2003).
\bibitem{Pfaffian}The square of the Pfaffian equals the determinant.
\bibitem{LiebWu68}E. H. Lieb and F. Y. Wu, Phys.~Rev.~Lett.~\textbf{20}, 1445 (1968).
\bibitem{FettesMorgenstern00}W. Fettes and I. Morgenstern, Computer Physics Communications \textbf{124},
148 (2000).
\bibitem{Gauge}P. Deuar and P. D. Drummond, Comp.~Phys.~Comm.\textbf{~142}, 442
(2001); P. Deuar and P. D. Drummond, Phys.~Rev.~A \textbf{66}, 033812
(2002); P.D. Drummond and P. Deuar, J. Opt.~B \textbf{5}, 281 (2003). 
\bibitem{TrivediCeperley90}N. Trivedi and D. M. Ceperley, Phys.~Rev.~B \textbf{41}, 4552 (1990).
\bibitem{Exact}S. Sorella, A. Parola, M. Parrinello and E. Tosatti, Europhys.~Lett.~\textbf{12},
721 (1990).\end{thebibliography}
\end{document}